\documentclass{aastex}
\usepackage{spr-astr-addons}
\usepackage{url}

\RequirePackage{color}

\begin{document}

\title{Theoretical Mass-Luminosity Relations in Gaia G-Band}
\slugcomment{Not to appear in Nonlearned J., 45.}
\shorttitle{Theoretical Mass-Luminosity Relations in Gaia G-Band}
\shortauthors{Malkov et al.}

\author{Oleg Malkov\altaffilmark{1}} \and \author{Dana Kovaleva\altaffilmark{1}}
\and
\author{Aleksandr Zhukov\altaffilmark{1,2,3}}
\and
\author{Olga Dluzhnevskaya\altaffilmark{1}}

\altaffiltext{1}{Institute of Astronomy, Russian Academy of Sciences, 48 Pyatnitskaya St., Moscow 119017, Russia}
\altaffiltext{2}{Joint Stock Company ``Special Research Bureau of Moscow Power Engineering Institute'', Moscow, Russia}
\altaffiltext{3}{Federal State Budgetary Scientific Institution ``Expert and Analytical Center'', Moscow, Russia}

\begin{abstract}
We have analyzed the retrieval of the relation between mass and absolute Gaia G magnitude as obtained from theoretical models by~\cite{2012MNRAS.427..127B} for the stars of different luminosity classes. For subgiant and main-sequence stars, we provide approximate analytical direct and inverse relations based on the most probable value of G magnitude in the course of evolution within the given stage. A comparison with (albeit few) observational data confirms that our
results can be used for the estimation of the stellar mass from Gaia photometry in the range of 1 to 10 solar mass. We argue that similar relations for other luminosity classes are not informative.
\end{abstract}

\keywords{stars: fundamental parameters, mass-luminosity relation}

\section{Introduction}
\label{sect:intro}

Knowledge of the mass of a star of given luminosity is often needed
to solve various problems of astrophysics.
In particular, in the course of our investigation
of orbital binaries \citep{2012A&A...546A..69M},
the necessity was felt to have general statistical relations between the
mass and the luminosity (determined from apparent magnitude, parallax and
interstellar extinction).
Such relations (hereafter, mass-luminosity relation, MLR) were constructed
and presented by different authors
\citep{1988BAICz..39..329H, 1991A&ARv...3...91A, 1993AJ....106..773H,
1999ApJ...512..864H, 2007MNRAS.382.1073M, 2010A&ARv..18...67T, 2021MNRAS.507.3583E}
for MS-stars for V and IR-bands.
To construct MLR for non-MS stars is a much more difficult problem
(see, e.g., \cite{1986A&A...168..161H, 1992msp..book.....S, 2020MNRAS.491.5489M}).
Note that the relationship between
the mass of a star and its absolute magnitude in one or another photometric band
is also traditionally called MLR.

Thanks to the success of the Gaia space mission \citep{2021A&A...649A...1G},
we have access to absolute G-photometry for $1.8 \cdot 10^9$ stars.
To properly process
these data one must have at one's disposal, in particular, MLR for G-band.
A detailed, precise observational G-band MLR for MS-stars can be found among
Mamajek's data available at
\url{http://www.pas.rochester.edu/~emamajek/EEM_dwarf_UBVIJHK_colors_Teff.txt}
and published in\\ \citep{2012ApJ...746..154P}, and \citep{2018MNRAS.479.5491E}.
However, there is no published G-band MLR for stars of other luminosity classes.

In the present paper we construct a subgiant G-band MLR.
In the solar vicinity,
there are about an order of magnitude fewer subgiants than MS-stars,
and the number of subgiants is comparable with the number of giants.
Thus, the construction of a subgiant G-band MLR seems to be an important and actual task.
This process is described in Section~\ref{sec:MLR-IV}.
Main sequence MLR is given in Section~\ref{sec:MLR-V}, while
MLR construction difficulties for other luminosity classes are discussed
in Section~\ref{sec:MLR-III}.
We draw our conclusions in Section~\ref{sec:conclusions}.

\section{Subgiants}
\label{sec:MLR-IV}

Stars enter the subgiant phase of evolution as soon as their core is depleted by hydrogen
 (less than 1\% by mass) and thermonuclear reactions stopped in the core.
 Fusion of hydrogen into helium continues in the envelope around the core, mainly by the CNO cycle.
In stars of mass less than about $0.4~M_\odot$
($0.34~M_\odot$, according to \cite{2018A&A...619A.177B},
 $0.45~M_\odot$, according to \cite{2018MNRAS.479.5491E})
this is impossible in principle
as they are fully convective and, therefore, chemically homogeneous,
 which means that when the core runs out of hydrogen, it also runs out in the whole star.
When stars of mass between 0.4 and $1.5~M_\odot$ complete the thermonuclear fusion in the core,
 they move into the shell hydrogen burning stage
 as the helium core of a uniform temperature (dT/dr = zero) grows gradually.

In more massive stars the energy release is more concentrated in the centre,
 so after the hydrogen in the core ends, the fusion in the core stops.
 Once it stops, the star shrinks until the conditions for shell hydrogen burning are reached,
 after which it moves to the subgiant branch.
 While the core is contracting, its temperature and luminosity are rising, it is moving up and to the right in the Hertzsprung-Russell diagram, and is passing through what is called a hook.

At the subgiant stage, the outer layers of the star expand and cool, the luminosity changes weakly, and the star moves to the right in the Hertzsprung-Russell diagram.
During this shell H burning stage, the star with $M > 1.5~M_\odot$
moves towards right with about same luminosity but $T_{\rm eff}$ is decreasing rapidly
until reaching the base of the red giant phase.
After that the star
will rapidly start to increase its size and luminosity, but its surface temperature
will barely change, and the star moves into the red giant branch
(see, e.g., \cite{2004gca..book.....K, 2013A&A...560A..16M}).

\begin{figure}                                                                                             
\centering
\includegraphics[width=8cm]{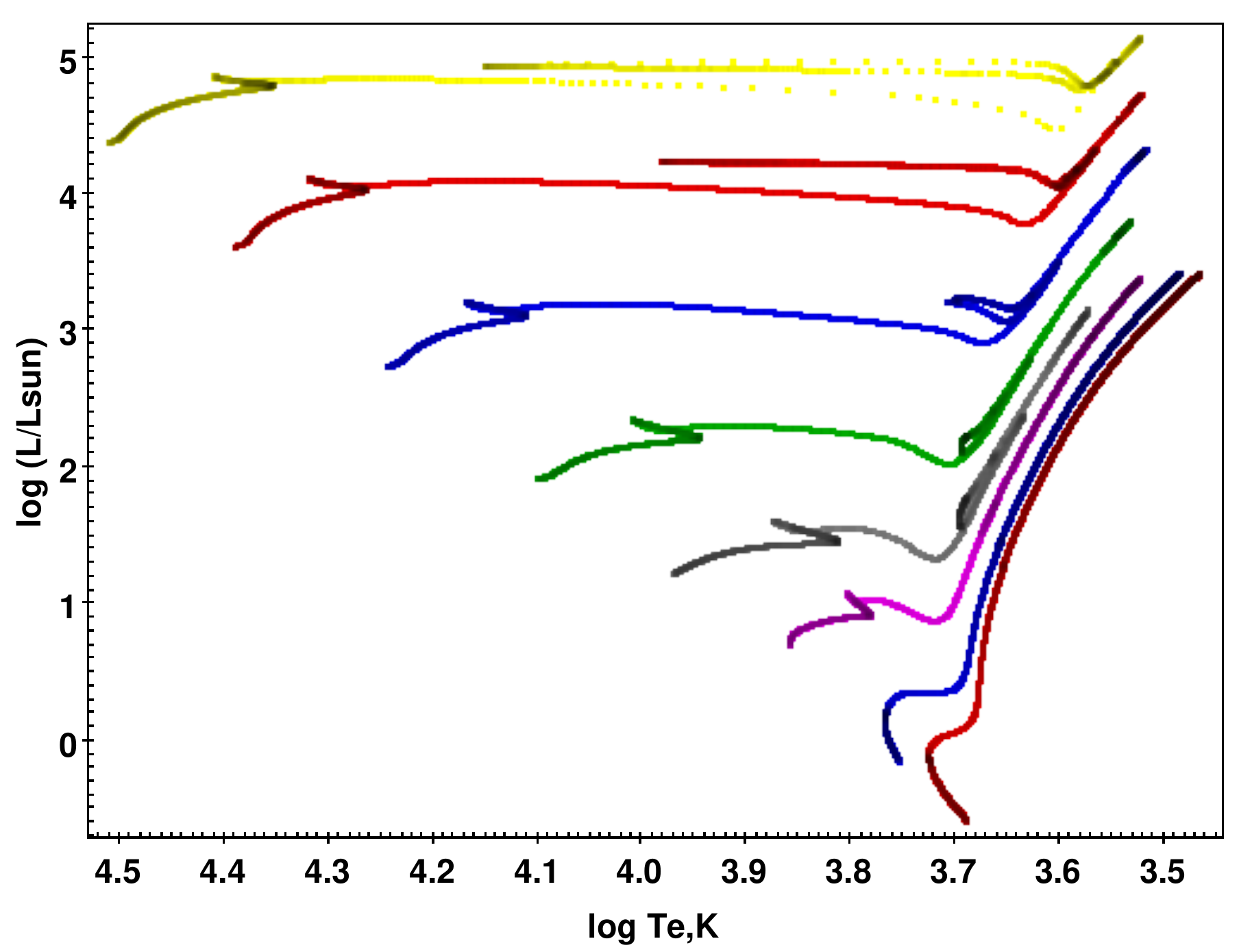}
\caption{Post-MS evolutionary tracks~\citep{2012MNRAS.427..127B}
for masses (from bottom to top) 0.8, 1, 1.5, 2, 3, 5, 9, 16 solar mass.
}
\label{fig:tracks}
\end{figure}

A typical system of evolutionary tracks for
intermediate mass stars (1-10, hereafter in solar mass) is shown in
Fig.~\ref{fig:tracks}.
To specify the subgiant region in the HRD
we consider the blue boundary of the subgiant stage to be the time of establishment of hydrogen shell source
(when the star starts to decrease its surface temperature after the hook),
and the red boundary to be the bottom of the red giant stage.

To construct a subgiant MLR we have used evolutionary tracks and isochrones
from \cite{2012MNRAS.427..127B}, with modifications, described in
\citep{2014MNRAS.444.2525C, 2015MNRAS.452.1068C, 2014MNRAS.445.4287T,
2017ApJ...835...77M, 2019MNRAS.485.5666P, 2020MNRAS.498.3283P} (see also 
\url{http://stev.oapd.inaf.it/cgi-bin/cmd}). For the Gaia photometric system, 
realisation described by \cite{2021A&A...649A...3R} is used.
Initial solar metallicity Z=0.0152 is accepted.
Using data of evolutionary tracks we assume that the actual mass of star
is equal to its initial mass. For our mass range (1-10) it seems to be
a reasonable approximation.

\begin{figure}
\centering
\includegraphics[width=7cm]{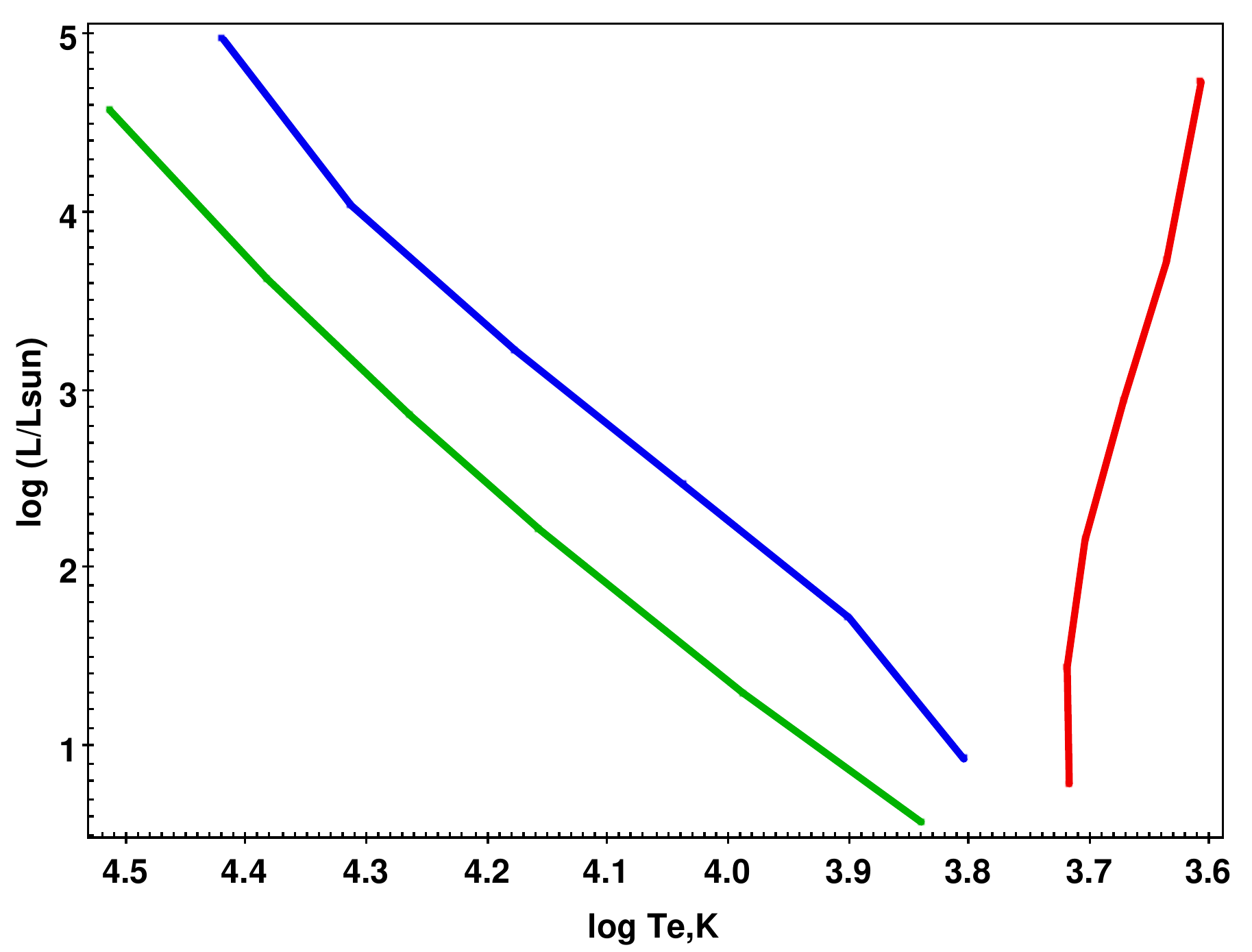}
\includegraphics[width=7cm]{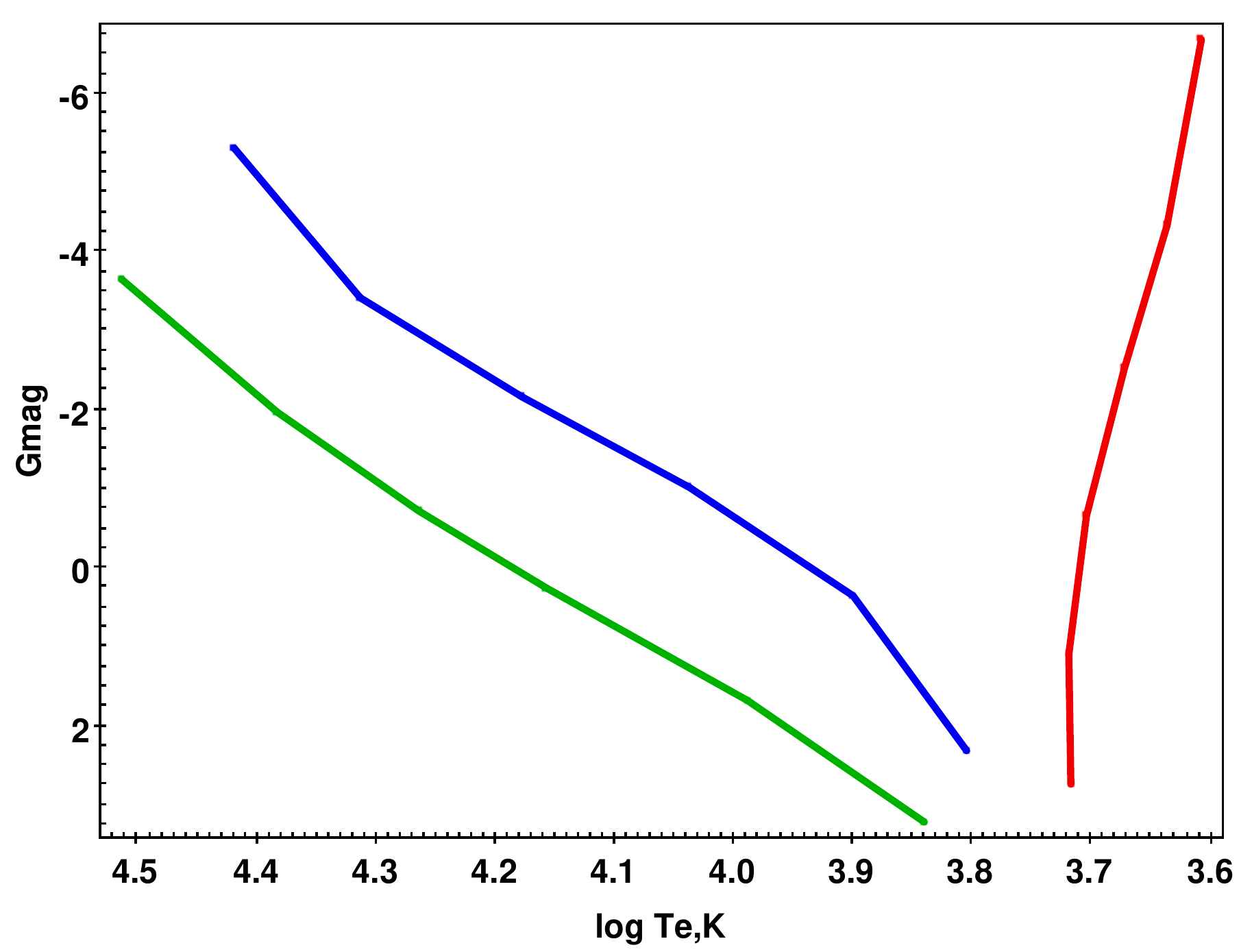}
\caption{Blue and red boundaries of the subgiant area, as well as ZAMS (green curve)
in the HRD on the logL (upper panel) and G-band (lower panel) scale.}
\label{fig:HRD}
\end{figure}

\begin{figure}
\centering
\includegraphics[width=7cm]{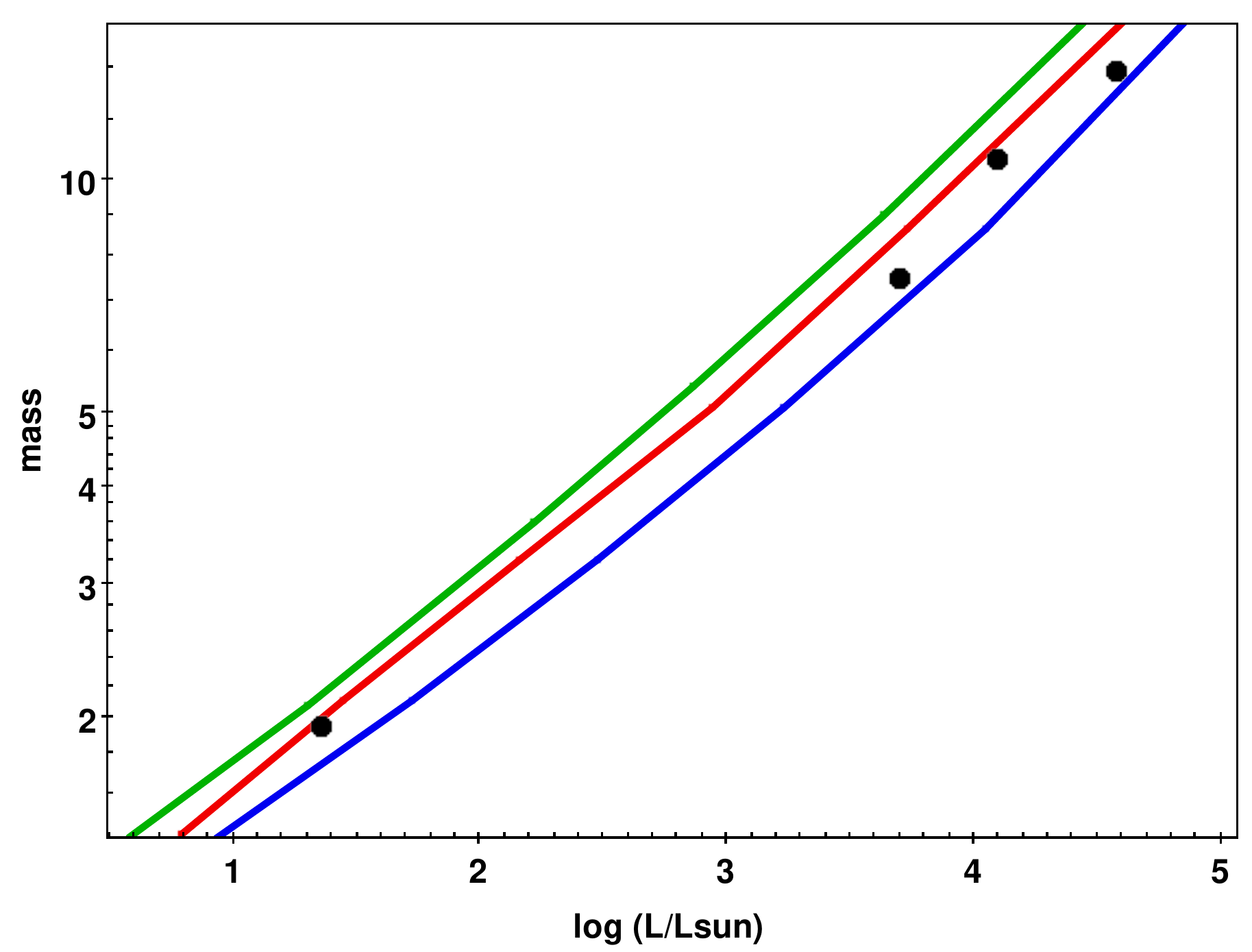}
\includegraphics[width=7cm]{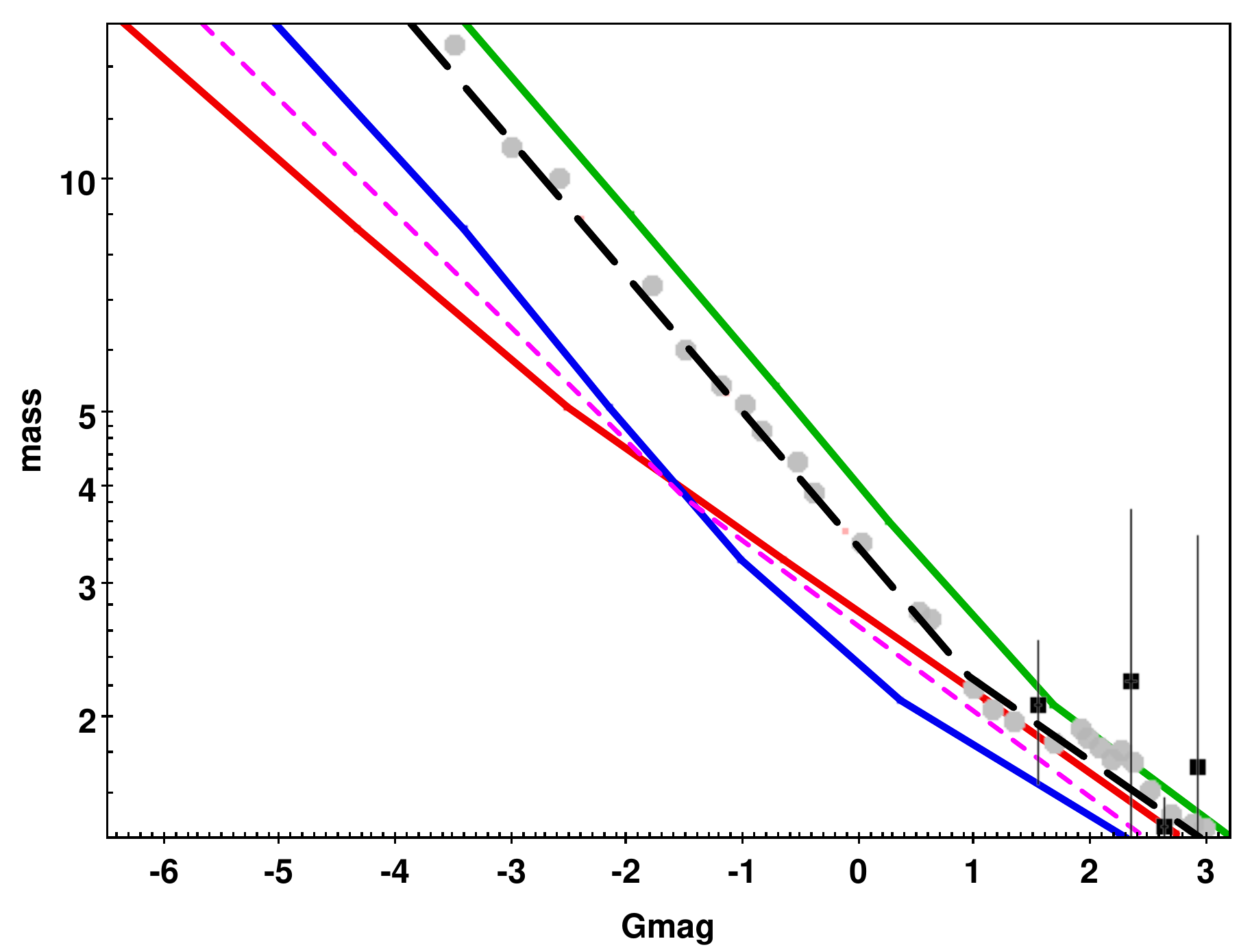}
\caption{Subgiant logL (upper panel) and absolute G-band (lower panel) mass-luminosity relations.                 
Mass is in solar units.
Red and blue curves are red and blue boundaries of the subgiant area, ZAMS is
represented by green curves.
Magenta dashed line is the linear approximation for subgiant MLR, see Eqs.~\eqref{equ:MLR1},~\eqref{equ:MLR2}.
Black dashed line is the linear approximation for MS MLR, see Eqs.~\eqref{equ:MLRV1},~\eqref{equ:MLRV2}.
Grey dots represent Mamajek's G-band MLR for MS-stars
(\url{http://www.pas.rochester.edu/~emamajek/EEM_dwarf_UBVIJHK_colors_Teff.txt}),
for comparison.
Black squares (lower panel) are components of ``twin'' subgiant
pairs from ORB6 with known G-photometry, see Table~\ref{tab:ORB6}.
Black circles (upper panel) are subgiant pairs from~\cite{2010A&ARv..18...67T} list, see Table~\ref{tab:Torres}
(here symbol sizes exceed observation errors).}
\label{fig:MLR}
\end{figure}

To construct a subgiant MLR we have selected starting and ending points of the subgiant phase on evolutionary tracks for our mass range.
The resulting subsequences are shown in Figs.~\ref{fig:HRD} (representing the Hertzsprung-Russell diagram)
and~\ref{fig:MLR} (representing mass-luminosity relation).
Blue and red curves indicate starting and ending points of the subgiant phase, respectively.
ZAMS (green curve) is also shown in Figs.~\ref{fig:HRD} and~\ref{fig:MLR}.

Fig.~\ref{fig:MLR} also contains the linear approximation for subgiant MLR (see Eqs. (1), (2))
and the linear approximation for MS MLR (see Section~\ref{sec:MLR-V} and Eqs. (3), (4)).
Mamajek's G-band MLR for MS-stars
(\url{http://www.pas.rochester.edu/~emamajek/EEM_dwarf_UBVIJHK_colors_Teff.txt})
is given for comparison.

Note that the G-band subgiant MLR is, on average, one or two magnitudes brighter than the MS MLR
(compare with 1-mag value, found by~\cite{1986A&A...168..161H} for V-band MS and subgiant MLR).
Note also an inversion in the G-band subgiant MLR:
the blue boundary is brighter than the red one for stars more massive than about 4,
and is fainter for less massive stars (Fig.~\ref{fig:MLR}, lower panel).

The subgiant G-band mass-luminosity relation can be drawn from Fig.~\ref{fig:MLR}.
Two considerations
allow us to assume that the mean value between the blue and red boundaries of the subgiant area
can be considered as a satisfactory approximation for the mean G-band subgiant MLR.
Firstly, at the subgiant stage the stellar luminosity changes relatively weakly (see Fig.~\ref{fig:tracks}).
Secondly, at this stage the star moves (to the right in the Hertzsprung-Russell diagram) at a nearly steady rate.
That mean subgiant G-band mass-luminosity relation can be linearly approximated and written as
\begin{equation}
M_G =
\begin{cases}
-6.74 \log m + 2.46 & \text{for $18>m>3.84$;}\\
-8.98 \log m + 3.77 & \text{for $3.84>m>1.4$}
\end{cases}
\label{equ:MLR1}
\end{equation}
and
\begin{equation}
\log m =
\begin{cases}
-0.15 M_G + 0.36 & \text{for $-6<M_G<-1.48$;}\\
-0.11 M_G + 0.42 & \text{for $-1.48<M_G<2.6$.}
\end{cases}
\label{equ:MLR2}
\end{equation}
Uncertainty depends on the mass range and does not exceed 0.4 mag and 0.1 $\log$ mass, respectively.
Correlation for both equations is not less than -0.996.

It is advisable to compare our results with empirical data.
Unfortunately, observational data on relatively precise stellar mass for subgiant binary systems
are rather scarce. Masses and luminosities for few subgiant binaries
from~\cite{2010A&ARv..18...67T} catalogue are listed in Table~\ref{tab:Torres} and
plotted in Fig.~\ref{fig:MLR} (upper panel, black circles).
They demonstrate a good agreement with our results.

As to the G-mag MLR, we attempt use of few subgiant binaries, selected from ORB6 catalogue~\citep{2001AJ....122.3472H}.
Dynamical mass of orbital binaries was calculated from their orbital elements (from ORB6)
and Gaia trigonometric parallaxes (results of ORB6-Gaia cross-matching are given in~\citep{2019INASR...3..360C}).
This procedure is described in~\cite{2012A&A...546A..69M}.
Absolute G-mag ($M_G$) was calculated from Gaia photometry and parallax,
and interstellar extinction (taken from Stilism service, see~\cite{2019A&A...625A.135L}).
We have selected pairs with high quality grades of orbital elements in ORB6 (grade = 1, 2, 3 or 8),
indication of presence of a subgiant in the spectral type classification,
and Gaia photometry available for both components.
Dynamical mass is calculated for the sum of components of binary system, so,
to obtain individual masses of components, we have selected pairs with a negligible
components' magnitudes difference (presumably ``twin'' or ``quasi-twin'' pairs).
The selected four systems are listed in Table~\ref{tab:ORB6} and
plotted in Fig.~\ref{fig:MLR} (lower panel, black squares).
One can see that these data are too scarce, inaccurate and local to make a definite conclusion
on a quality of our results.

\begin{table*}
\centering
\caption{Subgiant pairs from the list of~\cite{2010A&ARv..18...67T}}
\begin{tabular}{llcc}
\hline
Name          & Sp.Type   & mass                   & $\log$L           \\
\hline
V453 Cyg    A & B0.4IV    & 13.82   $\pm$  0.35    & 4.583 $\pm$ 0.026 \\
V453 Cyg    B & B0.7IV    & 10.64   $\pm$  0.22    & 4.094 $\pm$ 0.035 \\
V1388 Ori   A & B2.5IV-V  &  7.42   $\pm$  0.16    & 3.697 $\pm$ 0.044 \\
V1388 Ori   B & F7IV      &  1.945  $\pm$  0.027   & 1.361 $\pm$ 0.033 \\
\hline
\end{tabular}
\label{tab:Torres}
\end{table*}

\begin{table*}
\centering
\caption{Subgiant pairs from ORB6}
\begin{tabular}{llccr}
\hline
WDS        & Sp.Type    & system mass        & $M1_G$          & d$M_G$ \\
\hline                                                           
00093+7943 & A4IV       & 4.15 $\pm$ 0.44    & 1.50 $\pm$ 0.02 & 0.08 \\
12567-4741 & F5IV/V     & 4.47 $\pm$ 1.49    & 2.27 $\pm$ 0.03 & 0.15 \\
16044-1122 & F5IV+F5IV  & 2.90 $\pm$ 0.13    & 2.62 $\pm$ 0.01 & 0.02 \\
16564+6502 & F2IV       & 3.45 $\pm$ 2.56    & 2.85 $\pm$ 0.01 & 0.13 \\
\hline
\end{tabular}
\label{tab:ORB6}
\end{table*}

\section{Main sequence stars}
\label{sec:MLR-V}

It is advisable to verify our procedure and compare our results with available data.
Besides the boundaries for the G-band subgiant MLR, ZAMS line was allocated from evolutionary tracks
and drawn in Figs.~\ref{fig:HRD} and~\ref{fig:MLR}.

The rate of stellar evolution (steps in unit time) within the main-sequence band is not constant.
Further the star from the ZAMS, faster it moves to towards TAMS in the HRD.
Particularly, as our rough estimations show, median age (within the MS band) corresponds
to about 30\% of the way between ZAMS and TAMS (which is actually the blue boundary of the subgiant stage)
in the HRD. 
A corresponding MS G-band mass-luminosity relation can be extracted from Fig.~\ref{fig:MLR}
similarly to subgiant mass-luminosity relation (see Section~\ref{sec:MLR-IV} and
Eqs.~\eqref{equ:MLR1},~\eqref{equ:MLR2})
and written as
\begin{equation}
M_G =
\begin{cases}
-5.67 \log m + 2.96 & \text{for $18>m>2.27$;}\\
-9.53 \log m + 4.33 & \text{for $2.27>m>1.4$}
\end{cases}
\label{equ:MLRV1}
\end{equation}
and
\begin{equation}
\log m =
\begin{cases}
-0.18 M_G + 0.52 & \text{for $-5<M_G<0.94$;}\\
-0.10 M_G + 0.44 & \text{for $0.94<M_G<2.8$.}
\end{cases}
\label{equ:MLRV2}
\end{equation}
Uncertainty depends on the mass range and does not exceed 0.9 mag and 0.14 $\log$ mass, respectively.
Correlation for both equations is not less than -0.999.
Note that the luminosity of a star in the main sequence band
changes on a larger scale than in the subgiant area (compare the widths of MS and subgiant
bands in Fig.~\ref{fig:MLR}). 

Mamajek's MLR for G-band MS stars
is plotted for comparison (Fig.~\ref{fig:MLR}, lower panel, grey dots).
An excellent agreement between our and Mamajek's result demonstrates
the applicability of our method for construction of the mass-luminosity relations.

\section{Giants and other luminosity classes}
\label{sec:MLR-III}

As one can see in Fig.~\ref{fig:tracks}, the luminosity of giant stars varies,
in the evolutionary process, within a very wide range.
For example, the luminosity of a two solar mass star changes, from the bottom
to the top of the red giant branch, by two and a half orders of magnitude.
Moreover, luminosity change areas overlap for different masses.
All this leads to giant MLR becoming very broad and imprecise, and therefore meaningless.
A discussion about other relations for giant stars can be found, e.g., in~\cite{2020RAA....20..139M}.

A construction of MLR for other (I, II, VI, VII) luminosity classes is also quite complicated
(in particular, for white dwarfs the luminosity of a star of a given mass strongly depends
also on its age) and is not discussed in the present paper. Moreover,
representatives of these luminosity classes are relatively rare among ORB6 stars
with known MK spectral classification.

\section{Conclusions}
\label{sec:conclusions}

We have constructed subgiant mass-luminosity relation for G-band, based on theoretical models
by~\citep{2012MNRAS.427..127B} \url{http://stev.oapd.inaf.it/cgi-bin/cmd}. Suggestions on expected errors are provided.
Observational data for subgiants with known dynamical masses are scarce but demonstrate that our relations (see Eqs.\eqref{equ:MLR1} and \eqref{equ:MLR2})
can be used for estimation of subgiant mass from its G-mag.
We perform similar
procedure to obtain mass-luminosity relation of main sequence stars.

Comparing MS mass-luminosity relation in this study and the empirical relation given by Mamajek
 gave us confident that the mass-luminosity relations produced in this study are reliable
 within the error limits $\pm$0.9 mag and $\pm$0.14 log mass for the main sequence
 and $\pm$0.4 mag and $\pm$0.1 log mass for the subgiant mass-luminosity relations respectively.

\acknowledgments
We are grateful to our anonymous reviewer whose constructive comments greatly helped us to improve the paper.
We thank D.~Chulkov for his help in preparing the paper.
This research has made use of 
NASA's Astrophysics Data System.
This research made use of TOPCAT, an interactive graphical viewer and editor for tabular data \citep{2005ASPC..347...29T}.

\begin{authorcontribution}
The authors made equal contribution to this work; O.M. wrote the paper.
\end{authorcontribution}

\begin{fundinginformation}
This work has been partially supported by RFBR grant 19-07-01198 and NSFC/RFBR grant 20-52-53009.
\end{fundinginformation}

\begin{dataavailability}
The data underlying this article will be shared on reasonable request to the corresponding author.
\end{dataavailability}



\begin{ethics}
\begin{conflict}
The authors declare that they have no conflicts of interest.
\end{conflict}
\end{ethics}

\nocite{*}
\bibliographystyle{raa}
\bibliography{subgiant}

\end{document}